\documentclass[twocolumn,showpacs,preprintnumbers,amsmath,amssymb,superscriptaddress]{revtex4}

\usepackage{graphicx}% Include figure files
\usepackage{epstopdf}% Include figure files
\usepackage{dcolumn}% Align table columns on decimal point
\usepackage{bm}% bold math

\begin{document}

\preprint{ISS05-Proceedings}

\title{
$^{23}$Na NMR study of non-superconducting 
double-layer hydrate Na$_x$CoO$_2\cdot y$H$_2$O 
}

\author{Hiroto Ohta}
\email{shioshio@kuchem.kyoto-u.ac.jp}
\author{Yutaka Itoh} 
\author{Chishiro Michioka} 
\author{Kazuyoshi Yoshimura} 
\affiliation{Department of Chemistry, Graduate School of Science, Kyoto 
University, Kyoto 606-8502, Japan}

\date{\today}

\begin{abstract}
We report $^{23}$Na NMR studies of the polycrystalline samples of double-layer hydrated cobalt oxides Na$_x$CoO$_2\cdot y$H$_2$O ($x\sim$035 and $y\sim$~1.3) with the superconducting transition temperatures $T_{c}<$1.8 K and $\sim$4.5 K, and the dehydrated Na$_x$CoO$_2$ ($x\sim$0.35). The hyperfine field and the electric field gradient at the Na sites in the non-hydrated Na$_{0.7}$CoO$_2$ and the dehydrated Na$_{0.35}$CoO$_2$ are found to be significantly reduced by the hydration, which indicates a strong shielding effect of the intercalated water molecules on the Na sites. The temperature dependence of $^{23}$Na nuclear spin-lattice relaxation rate 1/$^{23}T_1$ of the non-superconducting double-layer hydrate Na$_x$CoO$_2\cdot y$H$_2$O is found to be similar to that of the non-hydrated Na$_{0.7}$CoO$_2$, whose spin dynamics is understood by $A$-type (intra-layer ferromagnetic and inter-layer antiferromagnetic) spin fluctuations. The superconducting phase is located close to the quantum critical point with the $A$-type magnetic instability.  
\end{abstract}


\maketitle
 
 \section{Introduction}
The double-layer hydrated cobalt oxide superconductor Na$_{0.35}$CoO$_2\cdot $1.3H$_2$O \cite{Takada} has attracted great interests in the itinerant electron magnetism on a triangular lattice. The spin frustration effect on a triangular lattice has been believed to realize the resonating valence bond, so-called RVB state \cite{PWAnderson,Baskaran} . At first, the effect of water intercalation was thought only to extend the interlayer distance between the CoO$_2$ layers. But now, it turns out that the diversity of the site occupation of Na ions \cite{Cava1,Cava2} and of H$_3$O$^{+}$ oxonium ions \cite{H3O}  makes the electronic phase rich. Although the intercalated water molecules are the key ingredients for occurrence of superconductivity, the electronic effects on the CoO$_2$ planes are poorly understood. 

	The parent non-hydrate Na$_{0.7}$CoO$_2$ is an itinerant electronic system with a strong thermoelectric power \cite{Terasaki}. The inelastic neutron scattering studies revealed the $A$-type spin fluctuations on the CoO$_2$ planes, i.e. intra-plane ferromagnetic and inter-plane antiferromagnetic instability at low temperatures \cite{Boothroyd}. The electrical resistivity shows characteristic of a magnetic quantum critical point \cite{Taillefer}. The $^{23}$Na nuclear spin-lattice relaxation time indicates the existence of two dimensional ferromagnetic spin fluctuations at low temperatures \cite{Ihara1} and suggests the motion of Na ions at high temperatures \cite{Gavilano}. The detailed $^{23}$Na NMR study revealed the vital role of charge degree of freedom \cite{Alloul}.

	We recently found successful synthesis methods of the non-superconducting double-layer hydrates with $T_c<$1.8 K and the derivative superconducting hydrates with $T_c\sim$4.5 K \cite{Ohta1}. In this paper, we report $^{23}$Na NMR studies for the non-superconducting, the derivative superconducting double-layer hydrates and the dehydrated ones. We observed that the Na nuclear spin-lattice relaxation rate 1/$^{23}T_1$ of the non-superconducting double-layer hydrate is of two order smaller than that of the parent Na$_{0.7}$CoO$_2$ but shows the similar temperature dependence. The electronic state on the CoO$_2$ planes of the non-superconducting double-layer hydrate is found to be nearly the same as that of the parent Na$_{0.7}$CoO$_2$, i.e. the $A$-type spin fluctuations at low temperatures. The intercalated water molecules act as a shielding effect on the Na sites from the CoO$_2$ planes. 

\begin{figure}[h]
 \begin{center}
 \includegraphics[width=1.0\linewidth]{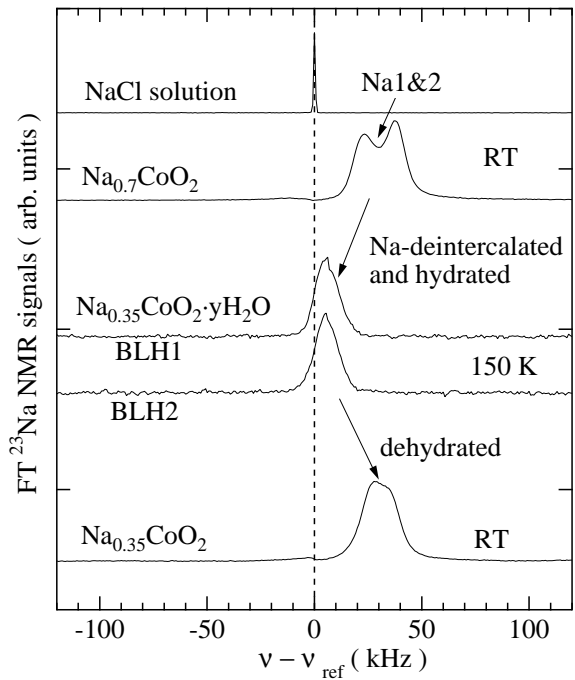}
 \end{center}
 \caption{\label{fig:NaNMR}
Fourier transformed $^{23}$Na NMR frequency spectra of a parent Na$_{0.7}$CoO$_2$, double-layer hydrated Na$_{0.35}$CoO$_2\cdot y$H$_2$O (BLH1 and BLH2), and the dehydrated Na$_{0.35}$CoO$_2$. The zero shift indicated by the dashed line is $\nu_{\mathrm{ref}}$=84.296 MHz ($\nu_{\mathrm{ref}}$=$^{23}\gamma_{\mathrm{n}}H$ with $^{23}\gamma_{\mathrm{n}}$=11.262 MHz/T and $H\sim$7.485 T), referred to the $^{23}$Na NMR line of NaCl in water at 300 K.    
}
 \end{figure}
 
 \section{Synthesis and experiments }
The parent compound Na$_{0.7}$CoO$_2$ was synthesized by a conventional solid-state reaction method. The powder of Na$_{0.7}$CoO$_2$ was immersed in Br$_2$/CH$_3$CN solution for 1 day to deintercalate Na$^{+}$ ions. The Na-deintercalated powder of Na$_{0.35}$CoO$_2$ was immersed in distilled water for 1 day to intercalate H$_2$O molecules. After filtering the powder of Na$_{0.35}$CoO$_2\cdot y$H$_2$O, it was put into a chamber with an atmosphere of 75 $\%$ humidity air. 
We found the duration (keeping time in the humidity-controlled chamber in a daily basis) dependence of the superconducting transition temperature $T_c$ and the physical properties of double-layer hydrates \cite{Ohta1}. Cryopreservation at -10 $^{\circ}$C was effective to quench further change of the each sample. The duration effect was sensitive to the initial condition of the samples and the atmosphere in the chamber. 

	Here, we prepared two samples of Na$_{0.35}$CoO$_2\cdot y$H$_2$O. The duration of the sample named BLH1 was 1 week, and that of BLH2 was 1 month. From the temperature dependence of the magnetic susceptibility at an external magnetic field of 20 Oe measured by a SQUID magnetometer, we estimated $T_c <$ 1.8 K and $\sim$4.5 K for BLH1 and BLH2, respectively. Since the samples with short duration did not show superconductivity down to 1.8 K, BLH1 is classified into these non-superconducting double-layer hydrates. Zero-field $^{59}$Co nuclear quadrupole resonance (NQR) experiments were performed to characterize these samples. The appearance of an internal magnetic field below about 4.5 K was confirmed in a broad $^{59}$Co NQR spectrum of BLH1, which is similar to that of the material in the previous report \cite{Ihara2}.  

	The $^{23}$Na (spin $I$=3/2) NMR experiments were performed by a coherent-type pulsed NMR spectrometer for the powdered samples. The $^{23}$Na NMR frequency spectra were obtained by a fast Fourier transformation technique for the free induction signals. The $^{23}$Na nuclear spin-lattice relaxation time $^{23}T_1$ was measured by an inversion recovery technique. The free induction signal $M(t)$ as a function of time $t$ after an inversion pulse and $M(\infty)[\equiv M(t>10T_1$)] were measured. The recovery curves $p(t)$$\equiv$1-$M(t)/M(\infty)$ were analyzed by $p(t)$=$p$(0)[0.1exp(-$t/T_1$)+0.9exp(-6$t/T_1$)] for central transition ($I_z$=1/2$\leftrightarrow$-1/2), where $p$(0) and $T_1$ are fit parameters.   

\begin{figure}[h]
 \begin{center}
 \includegraphics[width=1.05\linewidth]{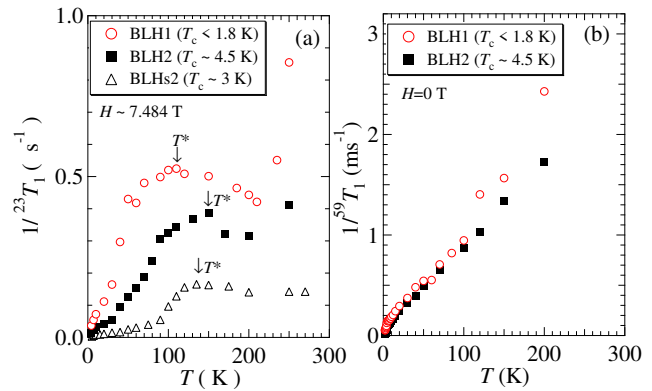}
 \end{center}
 \caption{\label{fig:T1NaCo}
(a) Temperature dependence of the $^{23}$Na nuclear spin-lattice relaxation rate 1/$^{23}T_1$ for BLH1, BLH2, and for BLHs2 ($T_c\sim$3 K) reproduced from Ref. \cite{Ohta2}. (b) Temperature dependence of the $^{59}$Co nuclear spin-lattice relaxation rate 1/$^{59}T_1$ for BLH1 and BLH2 by a zero field $^{39}$Co NQR technique.  
 }
 \end{figure}
  
 \section{NMR results and discussions}
	Figure 1 shows the Fourier transformed $^{23}$Na NMR spectra of the parent Na$_{0.7}$CoO$_2$, the double-layer hydrated BLH1 and BLH2, and the dehydrated Na$_{0.35}$CoO$_2$. The dehydrated Na$_{0.35}$CoO$_2$ was obtained by heating the double-layer hydrate powder at about 250 $^{\circ}$C and then putting immediately into a hydrophobic liquid \cite{Foo}.      
 
	In Na$_{x}$CoO$_2$, there are two crystallographic Na sites, Na1 and Na2, which locate just above the Co site and in the center of the Co triangles, respectively. The two $^{23}$Na NMR lines for Na$_{0.7}$CoO$_2$ in Fig. 1 are assigned to the central transition lines ($I_z$=1/2$\leftrightarrow$-1/2) of these Na1 and Na2 nuclei \cite{Alloul}. The satellite spectra are observed outside of Fig. 1. Both the Na 1 and Na2 NMR spectra are largely affected by Knight shifts and electric quadrupole shift. 
 
 	For the double-layer hydrates BLH1 and BLH2, however, the single NMR lines with small Knight shifts are observed. The satellite spectra seem to be wiped out. For the Na$_{0.35}$CoO$_2$ dehydrated from the double-layer hydrates Na$_{0.35}$CoO$_2\cdot y$H$_2$O, the largely shifted NMR line is observed once again. This spectrum of the dehydrate is different from that of the anhydrous in Ref. \cite{Alloul}. The recovery of Knight shift in the $^{23}$Na NMR spectra of the dehydrated Na$_{0.35}$CoO$_2$ demonstrates that not the Na deficiency but the intercalated water molecules diminish the Co-to-Na hyperfine field coupling and shield the electric field gradient from the CoO$_2$ planes.  

	Figure 2(a) shows the temperature dependence of the $^{23}$Na nuclear spin-lattice relaxation rate 1/$^{23}T_1$ for BLH1, BLH2 and BLHs2 ($T_c\sim$3 K) which is reproduced from Ref. \cite{Ohta2}. For comparison, the temperature dependence of the $^{59}$Co nuclear spin-lattice relaxation rate 1/$^{59}T_1$ for BLH1 and BLH2 are shown in Fig. 2(b). The magnitude of 1/$^{23}T_1$ significantly decreases from BLH1 ($T_c<$1.8 K) to BLH2 ($T_c \sim$4.5K), whereas that of 1/$^{59}T_1$ does not change so much. The broad maximum of 1/$^{23}T_1$, denoted by $T^{\ast}$, is seen in Fig. 2(a), whereas 1/$^{59}T_1$ is not in Fig. 2(b). We believe that the different $T_c$ results from difference in the degree of hydration and that the different shielding effect leads to the difference in spin dynamics at the Na site. 
	
\begin{figure}[h]
 \begin{center}
 \includegraphics[width=1.0\linewidth]{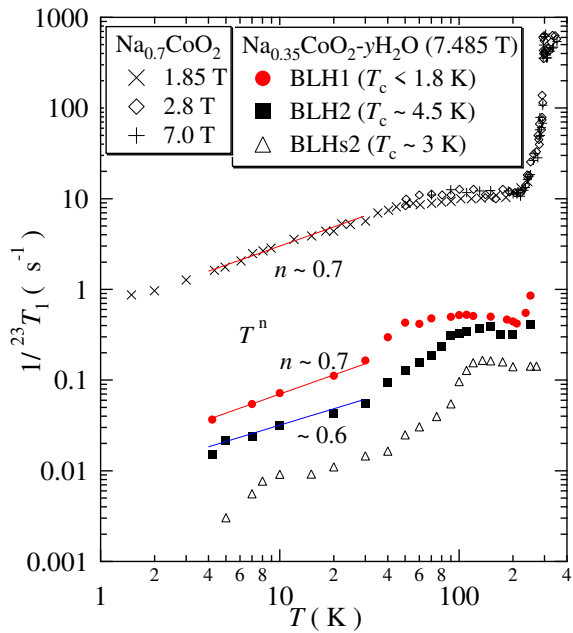}
 \end{center}
 \caption{\label{fig:T1Na}
$^{23}$Na nuclear spin-lattice rate 1/$^{23}T_1$ against temperature for Na$_{0.7}$CoO$_2$ in \cite{Ihara1,Gavilano}, the double-layer hydrates Na$_{0.35}$CoO$_2\cdot y$H$_2$O, BLH1, BLH2, and BLHs2 in \cite{Ohta2}. The solid lines are fits to the low temperature 1/$^{23}T_1$ by the power laws of $T^n$. 
 }
 \end{figure}

 \begin{figure}[h]
 \begin{center}
 \includegraphics[width=0.8\linewidth]{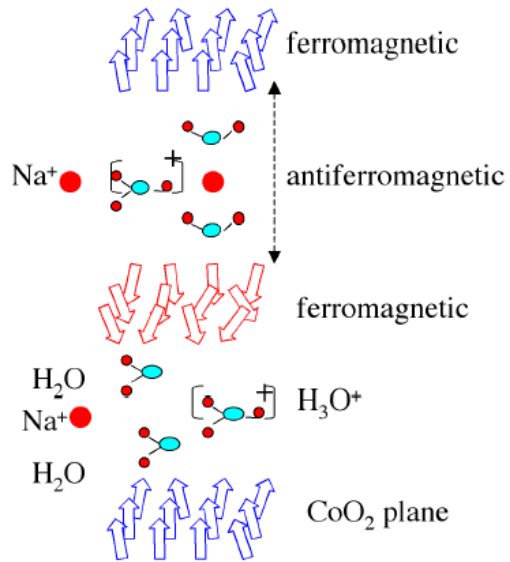}
 \end{center}
 \caption{\label{fig:Atype}
A snapshot of the $A$-type spin fluctuations for a double-layer hydrated system. 
The intra-plane spin fluctuations are ferromagnetic and the inter-plane spin fluctuations are antiferromagnetic.   
}
 \end{figure}
 
	The spin fluctuations probed by 1/$^{59}T_1$ of the in-plane $^{59}$Co nuclei are nearly independent of $T_c$, whereas the fluctuations probed by 1/$^{23}T_1$ of the inter-plane Na nuclei are strongly dependent on $T_c$. The different temperature dependence of 1/$^{23}T_1$ and 1/$^{59}T_1$ may result from the interplane spin fluctuations and/or the motion of Na ions. These interplane modes may control the value of $T_c$. 
	
	Figure 3 shows the log-log plots of 1/$^{23}T_1$ against temperature for the non-hydrated Na$_{0.7}$CoO$_2$ which are reproduced from Refs. \cite{Ihara1,Gavilano}, and for the double-layer hydrated BLH1, BLH2 and BLHs2 which is reproduced from Ref. \cite{Ohta2}. At low temperatures in the normal states above $T_c$, 1/$^{23}T_1$'s show the power-law behaviors (solid lines) of $T^n$ with $n\sim$0.7 and 0.6 for BLH1 and BLH2, respectively. These non-Korringa behaviors indicate that the samples are not conventional Fermi liquid systems. The magnitude of 1/$^{23}T_1$ significantly decreases from the non-hydrated to the double-layer hydrated samples. 
	
	The temperature dependence of 1/$^{23}T_1$ in BLH1 is nearly the same as that in the parent Na$_{0.7}$CoO$_2$. Both systems show the power law of $T^{0.7}$ in 1/$^{23}T_1$ at low temperatures below about 20 K and the rapid increase at high temperatures above about 200 K. The ratio of 1/$^{23}T_1$ of Na$_{0.7}$CoO$_2$ to that of BLH1 is $\sim$45 below 10 K, and in the case of BLH2 the ratio is $\sim$103 below 10 K. In general, 1/$^{23}T_1$ is expressed by $\propto$$A_{hf}^{2}S(\omega_n)$ with the hyperfine coupling constant $A_{hf}$ and a spin-fluctuation spectrum $S(\omega_n)$  at an NMR frequency 2$\pi \omega_n$ \cite{Moriya}. The difference in the magnitude of 1/$^{23}T_1$ can be ascribed to that in $A_{hf}$, whose ratio is estimated to be about 6.6 for BLH1 and about 10.1 for BLH2 below 10 K. Thus, the spin dynamics in $S(\omega_n)$ at the Na site in the double-layer hydrates is similar to that in the non-hydrated Na$_{0.7}$CoO$_2$. The $A$-type spin fluctuations, i.e. intra-plane ferromagnetic and inter-plane antiferromagnetic fluctuations, are established in the non-hydrated Na$_{0.7}$CoO$_2$ \cite{Boothroyd}. The intercalated water molecules may cut the inter-plane antiferromagnetic couplings. Nearly ferromagnetic intra-plane spin fluctuations persist in the double-layer hydrates. Figure 4 illustrates the $A$-type spin fluctuations for the double-layer hydrates. The arrows represent a snapshot of the spin fluctuations.  
  
 \section{Conclusion} 
The intercalated water molecules act as the shielding effect on the Co-to-Na magnetic and electric coupling. The spin fluctuations of the double-layer hydrated non-superconductor are similar to the $A$-type spin fluctuations of the non-hydrated Na$_{0.7}$CoO$_2$. The superconductivity appears close to the quantum critical point with the $A$-type magnetic instability. 

\section{acknowledgement} 
We thank Dr. T. Waki and Dr. M. Kato for their experimental supports, and Dr. K. Ishida for fruitful discussions. This study was supported by a Grant-in-Aid for Science Research on Priority Area, 'Invention of anomalous quantum materials' from the Ministry of Education, Science, Sports and Culture of Japan (Grant No. 16076210).

\end{document}